# Non-Markovian Caldeira—Leggett quantum master equation

A. O. Bolivar

Departamento de Física, Universidade Federal de Minas Gerais, Caixa Postal 702, 30123-970, Belo Horizonte, Minas Gerais, Brazil

**Abstract**

We obtain a non-Markovian quantum master equation directly from the quantization of a non-Markovian Fokker—Planck equation describing the Brownian motion of a particle immersed in a generic environment (e.g. a non-thermal fluid). As far as the especial case of a heat bath comprising of quantum harmonic oscillators is concerned, we derive a non-Markovian Caldeira—Leggett master equation on the basis of which we work out the concept of non-equilibrium quantum thermal force exerted by the harmonic heat bath upon the Brownian motion of free particle. The classical limit (or dequantization process) of this sort of non-equilibrium quantum effect is scrutinized, as well.

## I. INTRODUCTION

From the mathematical point of view, the dynamics of an isolated material point is deterministically described by the Schrödinger equation

$$\frac{d\Psi(t)}{dt} = \mathcal{O}\Psi(t), \tag{1}$$

where the operator $\mathcal{O}$ acting upon the time-dependent function $\Psi(t)$ can carry features inherent in the particle such as its mass $m$, its position $x$ or its momentum $p$. Moreover, the mathematical structure of $\mathcal{O}$ also relies on the phenomenological parameter $\hbar$, dubbed the Planck constant $h$ divided by $2\pi$, which is responsible for the signature of the quantum world. It is worth stressing that in the Schrödinger equation the quantities $m$, $t$, as well as $x$ or $p$ show up as classical physical quantities in the sense that they are $\hbar$-independent. Most especially, the Schrödinger functions $\Psi(t)$ account for the remarkable phenomenon of superposition or interference of quantum states which is in the core of current researches on quantum computation [1].

Yet, from the physical standpoint, a quantum system cannot be imagined as being in isolation from its surroundings. In truth, it only comes into existence as far as its interaction with a certain environment (e.g., a measuring apparatus) is concerned. Accordingly, a quantum system must indeed be idealized as an open quantum system comprising of a tagged particle immersed in a generic quantum environment undergoing a jittering movement dubbed quantum Brownian motion that in turn is to be mathematically described by master equations of the general form

$$\frac{d\rho(t)}{dt} = \mathcal{L}\rho(t), \tag{2}$$

where the superoperator $\mathcal{L}$ (the so-termed Liouvillian of the quantum open system) acting on the von Neumann density operator $\rho(t)$ bears some environmental features, such as coupling constants (a sort of friction constant) and the fluctuation energy accounting for the existence of the quantum Brownian movement, as well as some properties inherent in the tagged particle such as its mass $m$, its position $x$, and the Planck constant $\hbar$. In theory of quantum open systems the pivotal issue is therefore the following [2]: *How can we build up or derive some physically meaningful Liouvillians $\mathcal{L}$?*

From the mathematical viewpoint, the superoperator $\mathcal{L}$ can be algebraically built up as Lindblad operators by beginning with a system-environment model Hamiltonian, and then making the Born and Markov approximations. This approach describes a sort of Markovian interaction between the open system (e.g., the Brownian particle) and its environment. In addition, it is assumed that the system and environment begin in a product state, i.e., they are initially independent and non-interacting [3—6]. Applications of the Lindblad formalism can be found, for instance, in quantum optics in which the environment is represented by a quantized radiation field while the Brownian particle is deemed to be an atom or a molecule. The corresponding master equation is called quantum optical master equation [5,6].

On the other hand, in the context of quantum information theory the study of the Lindblad quantum master equations seems to be of vital importance for physical situations in which the coupling between particle and environment is deemed to be too weak in order that the coherence property of the Schrödinger functions can be preserved in time, as long as the decoherence process, i.e., the destruction of the superposition states by the inevitable presence of environment, can be meticulously controlled [1,5]. Nevertheless, one has been argued that "*one should not attribute fundamental significance to the Lindblad master equation*" because "*the Lindblad theory is not applicable in most problems of solid state physics at low temperatures for which neither the Born approximation is valid nor the Markov assumption holds*" [7]. Furthermore, recent controversies [8] seem to point out the inadequacy of the Lindblad approach to fathoming the true physics of open quantum systems [4].

Alternatively, on the ground of the path-integral approach to open systems, Caldeira, Leggett, Cerdeira, and Ramaswamy derived two Markovian quantum master equations, the so-called Caldeira—Leggett equations (CLEs) [4,5,9,10], in order to describe the quantum Brownian motion on the basis of assumptions reliant on the Hamiltonian modeling the environment as a bath of quantum harmonic oscillators. The first Markovian CLE [9] holds valid at high temperatures for any friction constant whereas the second one [10] for any temperature and very weak damping. Nevertheless, such Markovian CLEs may give rise to unphysical results, for they are not of the Lindblad form[3—6, 12], albeit the high-temperature Caldeira—Leggett equation [9] has been employed for looking at the decoherence phenomenon [5,6,11]. In brief, it has been claimed that Markovian CLEs cannot be considered as a *bona fide* description of quantum Brownian motion [7,12].

In order to contribute to a general theory of quantum open systems, in this article we set out to tackle the problem of deriving a non-Markovian Caldeira—Legget master equation, thereby eschewing such complaints against the applicability range of this class of non-Lindblad quantum master equations. More specifically, on the basis of our non-Markovian CLE we predict the existence of a non-equilibrium quantum thermal force exerted by the harmonic heat bath upon the Brownian motion of a free particle.

So, the present paper is laid out as follows. In Sect. II we derive a non-Markovian Klein—Kramers equation. Then, we obtain in the Sect. III the non-Markovian Caldeira—Leggett equation by means of the so-termed dynamical quantization [4,13]. In Sect. IV, on the basis of our Caldeira—Leggett master equation we evaluate the thermal quantum force exerted by the heat bath on the Brownian motion of a free particle. The classical limit of this sort of quantum force is examined, as well. Concluding remarks are reckoned with in Sect. V. Further, four appendices are attached.

## II. THE NON-MARKOVIAN KLEIN—KRAMERS EQUATION

The erratic motion of a tagged material point immersed in a generic environment (a paradigmatic example of open system) may be described by a set of stochastic differential equations (the so-called Langevin equations) [4,14]

$$\frac{dP}{dt} = -\frac{dV(X)}{dX} - 2\gamma P + b\Psi(t), \qquad (3)$$

$$\frac{dX}{dt} = \frac{P}{m}, \qquad (4)$$

where the environmental force,

$$F_{\text{env}}(P, \Psi) = -2\gamma P + b\Psi(t), \qquad (5)$$

is made up by the *memoryless* dissipative force $F_d = -2\gamma P$, which accounts for stopping the particle's motion via the dissipation coefficient $2\gamma \geq 0$, as well as by the Langevin force, $L(t) = b\Psi(t)$, which is responsible for activating the particle's movement through fluctuations the strength of which is measured by the parameter $b \geq 0$. Further, we can readily check that the parameter $b$ may be expressed in dimensions of $[mass \times length \times time^{-3/2}]$, since the function $\Psi(t)$ is in dimensions of $[time^{-1/2}]$.

From the mathematical viewpoint, in the Langevin equations (3) and (4) the time $t$, the friction constant $\gamma$, the mass $m$, and the fluctuation strength $b$ are deemed to be deterministic parameters, whereas the time-dependent quantities $X = X(t)$, $P = P(t)$, and $\Psi = \Psi(t)$ are viewed as random variables in the sense that there exists a probability distribution function, $\mathcal{F}_{XP\Psi}(x, p, \psi, t)$, associated with the stochastic system $\{X, P, \Psi\}$, expressed in terms of the possible values $x = \{x_i(t)\}$, $p = \{p_i(t)\}$, and $\psi = \{\psi_i(t)\}$, with $i \geq 1$, distributed about the sharp values $q$, $p'$ and $\varphi$ of $X$, $P$, and $\Psi$, respectively. It is assumed that the average value of any physical quantity $A(X, P, \Psi, t)$ can be calculated as

$$\langle A(X, P, \Psi, t) \rangle = \iiint_{-\infty}^{+\infty} a(x, p, \psi, t) \mathcal{F}_{XP\Psi}(x, p, \psi, t) dx dp d\psi, \qquad (6)$$

fulfilling the normalization condition

$$\langle 1 \rangle = \iiint\limits_{-\infty}^{+\infty} \mathcal{F}_{XP\Psi}(x,p,\psi,t)dxdpd\psi = 1. \tag{7}$$

The set of stochastic differential equations (3) and (4) gives rise to the following Fokker—Planck equation in the Gaussian approximation (see Appendix A)

$$\frac{\partial \mathcal{F}}{\partial t} = -\frac{p}{m}\frac{\partial \mathcal{F}}{\partial x} + \frac{\partial}{\partial p}\left[\frac{\partial \mathcal{V}_{\text{eff}}(x,t)}{\partial x} + 2\gamma p\right]\mathcal{F} + \mathcal{D}(t)\frac{\partial^2 \mathcal{F}}{\partial p^2} \tag{8}$$

for the (marginal) probability distribution function $\mathcal{F} \equiv \mathcal{F}(x,p,t) = \int_{-\infty}^{\infty} \mathcal{F}_{XP\Psi}(x,p,\psi,t)d\psi$. Equation (8) is valid in the Gaussian approximation since it is expressed in terms of the Gaussian properties of the Langevin force in (3), namely, its mean value $\langle L(t) \rangle$ as well as its autocorrelation function $\langle L(t)L(t') \rangle$. More specifically, $\langle L(t) \rangle = b\langle \Psi(t) \rangle$, with

$$\langle \Psi(t) \rangle = \lim_{\varepsilon \to 0} \frac{1}{\varepsilon} \int_{t}^{t+\varepsilon} \langle \Psi(t') \rangle dt', \tag{9}$$

gives rise in (8) to the effective potential

$$\mathcal{V}_{\text{eff}}(x,t) = V(x) - xb\langle \Psi(t) \rangle, \tag{10}$$

while $\langle L(t)L(t') \rangle$ generates the time-dependent diffusion coefficient

$$\mathcal{D}(t) = 2\gamma m \mathcal{E}(t) \tag{11}$$

that in turn is expressed in terms of the following function

$$\mathcal{E}(t) = \frac{1}{4\gamma m} \lim_{\varepsilon \to 0} \frac{1}{\varepsilon} \iint_{t}^{t+\varepsilon} \langle L(t')L(t'') \rangle dt' dt'' = \frac{b^2}{4\gamma m} I(t), \tag{12}$$

with the dimensionless time-dependent function $I(t)$ given by

$$I(t) = \lim_{\varepsilon \to 0} \frac{1}{\varepsilon} \iint_{t}^{t+\varepsilon} \langle \Psi(t')\Psi(t'') \rangle dt' dt''. \tag{13}$$

It is readily to check that the time-dependent function $\mathcal{E}(t)$ has dimensions of energy, i.e., $[mass \times length^2 \times time^{-2}]$. Hence we call it the diffusion energy responsible for the Brownian motion of the particle immersed in a generic environment.

In order to determine a relationship between the fluctuation parameter $b$ and the dissipation parameter $\gamma$ present in (3) and (8), we assume that the solution $\mathcal{F}(x,p,t)$ to (8) renders steady in the long-time limit $t \to \infty$, i.e., $\lim_{t\to\infty}\mathcal{F}(x,p,t) \approx \mathcal{F}(x,p)$, with the effective potential (10) being given by $\lim_{t\to\infty}\mathcal{V}_{\text{eff}}(x,t) \approx V(x) - xb\langle\Psi(\infty)\rangle$, and the diffusion energy (12) displaying the following asymptotic behavior

$$\lim_{t\to\infty}\mathcal{E}(t) \approx \mathcal{E}(\infty) = \frac{b^2}{4\gamma m}, \qquad (14)$$

provided that

$$\lim_{t\to\infty} I(t) \approx I(\infty) = 1. \qquad (15)$$

The physical significance of condition (15) has to do with the fact that environmental fluctuations do possess Markovian correlations, i.e., the correlational function $I(t)$ displays a local (short) range behavior decaying to one in the steady regime. By contrast, non-Markovian effects show up in the nonequilibrium regime $0 < t < \infty$. Moreover, condition (14) yields the Markovian fluctuation-dissipation relation in the form

$$b = \sqrt{4\gamma m \mathcal{E}(\infty)}. \qquad (18)$$

The parameter $b$ measuring the strength of the fluctuations is therefore determined in terms of the friction constant $\gamma$, the mass of the particle $m$, as well as the steady diffusion energy $\mathcal{E}(\infty)$.

The characteristic feature underlying the concept of time-dependent diffusion energy (12), $\mathcal{E}(t) = \mathcal{E}(\infty)I(t)$, is that it sets up a general relationship between fluctuation and dissipation processes as well as fulfilling the validity condition $0 < \mathcal{E}(t) < \infty$. Both cases $\mathcal{E}(t) = 0$ and $\mathcal{E}(t) = \infty$ should be disregarded, for they may violate the fluctuation—dissipation relation. The former case may lead to dissipation without fluctuation, while the latter one may give rise to fluctuation without dissipation.

Summarizing, the Brownian motion of a particle immersed in a generic stationary environment is described by the Fokker—Planck equation

$$\frac{\partial\mathcal{F}}{\partial t} = -\frac{p}{m}\frac{\partial\mathcal{F}}{\partial x} + \frac{\partial}{\partial p}\left[\frac{\partial\mathcal{V}_{\text{eff}}(x,t)}{\partial x} + 2\gamma p\right]\mathcal{F} + 2\gamma m\mathcal{E}(\infty)I(t)\frac{\partial^2\mathcal{F}}{\partial p^2}, \qquad (19)$$

starting from the deterministic initial condition $\mathcal{F}(x,p,t=0) = \delta(x)\delta(p)$ and evolving towards a steady solution. The effective potential $\mathcal{V}_{\text{eff}}(x,t)$ is given by

$$\mathcal{V}_{\text{eff}}(x,t) = V(x) - x\sqrt{4\gamma m\mathcal{E}(\infty)}\langle\Psi(t)\rangle. \qquad (20)$$

The Fokker—Planck equation (19) is generated by the Langevin equations

$$\frac{dP}{dt} = -\frac{dV(X)}{dX} - 2\gamma P + \sqrt{4\gamma m\mathcal{E}(\infty)}\Psi(t),$$

$$\frac{dX}{dt} = \frac{P}{m}.$$

On the condition that $\langle\Psi(t)\rangle = 0$, $I(t) = 1$, and the environment is in thermodynamic equilibrium at temperature $T$, so that we can identify the diffusion energy $\mathcal{E}(\infty)$ with the thermal energy $k_B T$, where $k_B$ is the Boltzmann constant, then the non-Markovian Fokker—Planck equation (19) reduces to the Markovian Klein—Kramers equation [4,14,15]:

$$\frac{\partial\mathcal{F}}{\partial t} = -\frac{p}{m}\frac{\partial\mathcal{F}}{\partial x} + \frac{\partial}{\partial p}\left[\frac{dV(x)}{dx} - 2\gamma p\right]\mathcal{F} + 2\gamma m k_B T\frac{\partial^2\mathcal{F}}{\partial p^2}. \tag{21}$$

We wish to point out that the Klein—Kramers equation (21) has been derived without postulating *ab initio* the statistical properties of the Langevin force $L(t)$

$$\langle L(t)\rangle = 0, \tag{22}$$

$$\langle L(t)L(t')\rangle = 4\gamma m k_B T\delta(t-t'), \tag{23}$$

as is commonly made in the literature [14,15]. In our approach the important fact is the asymptotic behavior of $\langle\Psi(t)\rangle$ and $I(t)$ in the stationary limit. In other words, both the statistical properties (22) and (23) turn up as sufficient but not necessary conditions for attaining the steady state.

Lastly, because our Fokker—Planck equation (19) contains the Markovian Klein—Kramers equation (21) as an especial case, we dub it the non-Markovian Klein—Kramers equation.

### III. THE DYNAMICAL QUANTIZATION

Having derived the non-Markovian Klein—Kramers equation (19) for a Brownian particle immersed in a generic Gaussian environment, we now wish to quantize it by means of the dynamical quantization process [4,13]. First, we introduce the following Fourier transform

$$\chi(x,\eta,t) = \int_{-\infty}^{\infty}\mathcal{F}(x,p,t)e^{ip\eta}dp, \tag{24}$$

such that the exponential $e^{ip\eta}$ is a dimensionless term. Upon inserting (24) into our non-Markovian Fokker—Planck equation (19), we obtain the classical equation of motion in space $(x,\eta)$

$$\frac{\partial \chi}{\partial t} = -i\eta \frac{\partial \mathcal{V}_{\text{eff}}(x,t)}{\partial x} \chi + \frac{i}{m} \frac{\partial^2 \chi}{\partial x \partial \eta} - 2\gamma\eta \frac{\partial \chi}{\partial \eta} - 2\gamma m \mathcal{E}(\infty) I(t) \eta^2 \chi, \tag{25}$$

where $\chi \equiv \chi(x, \eta, t)$.

The stochastic dynamics (19) is said to be quantized by introducing into the equation of motion (25) the quantization conditions through the change of variables $(x, \eta) \mapsto (x_1, x_2)$ given by

$$x_1 = x + \frac{\eta \hbar}{2} \tag{26}$$

and

$$x_2 = x - \frac{\eta \hbar}{2}, \tag{27}$$

whereby the transformation parameter $\hbar$ having dimensions of angular momentum, i.e., $[mass \times length^2 \times time^{-1}]$, is dubbed Planck's constant that in turn is responsible for the signature of quantum realm.

The geometric meaning of the quantization conditions (26) and (27) lies at the existence of a minimal distance between the points $x_1$ and $x_2$ due to the quantum nature of space, i.e., $x_2 - x_1 = \eta \hbar$, such that in the classical limit $\hbar \to 0$, physically interpreted as $|\eta \hbar| \ll |x_2 - x_1|$, the result $x_2 = x_1 = x$ can be readily recovered.

Upon making use of the relations

$$\frac{\partial}{\partial \eta} = \frac{\partial x_1}{\partial \eta} \frac{\partial}{\partial x_1} + \frac{\partial x_2}{\partial \eta} \frac{\partial}{\partial x_2} = \frac{\hbar}{2} \left( \frac{\partial}{\partial x_1} - \frac{\partial}{\partial x_2} \right),$$

$$\frac{\partial}{\partial x} = \frac{\partial x_1}{\partial x} \frac{\partial}{\partial x_1} + \frac{\partial x_2}{\partial x} \frac{\partial}{\partial x_2} = \frac{\partial}{\partial x_1} + \frac{\partial}{\partial x_2},$$

as well as the Gaussian approximation in the quantum context (see Appendix B)

$$\mathcal{V}_{\text{eff}}^{(\hbar)}(x_1, t) - \mathcal{V}_{\text{eff}}^{(\hbar)}(x_2, t) \sim \eta \hbar \frac{\partial \mathcal{V}_{\text{eff}}^{(\hbar)}(x,t)}{\partial x}, \tag{28}$$

we obtain the non-Markovian quantum master equation

$$i\hbar \frac{\partial \rho}{\partial t} = \left[ \mathcal{V}_{\text{eff}}^{(\hbar)}(x_1, t) - \mathcal{V}_{\text{eff}}^{(\hbar)}(x_2, t) \right] \rho - \frac{\hbar^2}{2m} \left( \frac{\partial^2 \rho}{\partial x_1^2} - \frac{\partial^2 \rho}{\partial x_2^2} \right) - i\hbar\gamma (x_1 - x_2) \left( \frac{\partial \rho}{\partial x_1} - \frac{\partial \rho}{\partial x_2} \right)$$
$$- \frac{i \mathcal{D}_\hbar(t)}{\hbar} (x_1 - x_2)^2 \rho \tag{29}$$

which describes the quantum Brownian motion of a particle moving in the effective potentials

$$\mathcal{V}_{\text{eff}}^{(\hbar)}(x_1, t) = V(x_1) - x_1\sqrt{4\gamma m \mathcal{E}_\hbar(\infty)}\langle\Psi(t)\rangle \tag{30}$$

and

$$\mathcal{V}_{\text{eff}}^{(\hbar)}(x_2, t) = V(x_2) - x_2\sqrt{4\gamma m \mathcal{E}_\hbar(\infty)}\langle\Psi(t)\rangle. \tag{31}$$

Our master equation (29) complies with the quantum fluctuation-dissipation relationship given by the quantum diffusion coefficient

$$\mathcal{D}_\hbar(t) = 2\gamma m \mathcal{E}_\hbar(t). \tag{32}$$

Upon performing the transition from the classical equation of motion (25) to the quantum dynamics (29) via quantization conditions (26) and (27), we have replaced the solution $\chi \equiv \chi(x, \eta, t)$ with $\rho \equiv \rho(x_1, x_2, t)$, which now turns out to depend on the Planck constant, $\hbar$. It is relevant to notice that in the quantum master equation (29), the time evolution parameter $t$, the mass $m$, the frictional constant $\gamma$, as well as the both functions $\langle\Psi(t)\rangle$ and $I(t)$ are deemed to be non-quantized quantities, that is, they are $\hbar$-independent, whereas the classical diffusion energy $\mathcal{E}(\infty)$ has been subject to a quantization process, i.e., $\mathcal{E}(\infty) \to \mathcal{E}_\hbar(\infty)$.

Equation (29) describes the quantum Brownian motion of a particle immersed a generic Gaussian quantum environment. For the specific case of a heat bath comprising of quantum harmonic oscillators with oscillation frequency $\omega$, if we could identify the Brownian particle's steady diffusion energy $\mathcal{E}_\hbar(\infty)$ with the mean thermal energy of the bath in thermodynamic equilibrium (see Appendix C), then we obtain

$$\mathcal{E}_\hbar(\infty) = \frac{b_\hbar^2}{4\gamma m} = \frac{\omega\hbar}{2}\coth\left(\frac{\omega\hbar}{2k_B T}\right). \tag{33}$$

Accordingly, the equilibrium quantum fluctuation—dissipation relation reads

$$b_\hbar = \sqrt{2\gamma m \omega\hbar \coth\left(\frac{\omega\hbar}{2k_B T}\right)} \tag{34}$$

reducing to (18) in the classical limit $k_B T \gg \omega\hbar/2$. Moreover, the quantum diffusion constant associated with the thermal diffusion energy (33) reads

$$\mathcal{D}_\hbar(\infty) = \gamma m \omega\hbar \coth\left(\frac{\omega\hbar}{2k_B T}\right), \tag{35}$$

yielding at high temperatures, $T \gg \omega\hbar/2k_B$, the classical diffusion constant in (21): $\mathcal{D}(\infty) = 2\gamma m k_B T$. On the other hand, the zero-point diffusion constant reads $\mathcal{D}_\hbar^{(T=0)}(\infty) = \gamma m \omega\hbar$. Thus upon making use of (35) in (29) we obtain the non-Markovian quantum master equation

$$i\hbar \frac{\partial \rho}{\partial t} = [\mathcal{V}_{\text{eff}}(x_1, t) - \mathcal{V}_{\text{eff}}(x_2, t)] \rho - \frac{\hbar^2}{2m} \left( \frac{\partial^2 \rho}{\partial x_1^2} - \frac{\partial^2 \rho}{\partial x_2^2} \right)$$

$$-i\hbar\gamma(x_1 - x_2)\left(\frac{\partial \rho}{\partial x_1} - \frac{\partial \rho}{\partial x_2}\right) - i\gamma m\omega \coth\left(\frac{\omega\hbar}{2k_B T}\right) I(t)(x_1 - x_2)^2 \rho \qquad (36)$$

which describes a Brownian particle immersed in a quantum heat bath subject to averaging and non-Markovian correlational effects through the effective potentials $\mathcal{V}_{\text{eff}}(x_k, t)$, with $k = 1,2$, and the function $I(t)$, respectively. It is worth underscoring that Eq.(36) has been derived for any initial condition $\rho(x_1, x_2, t = 0)$.

For $\langle \Psi(t) \rangle = 0$ and $I(t) = 1$, our quantum master equation (36) reduces to the Markovian Caldeira—Leggett equation

$$i\hbar \frac{\partial \rho}{\partial t} = [V(x_1, t) - V(x_2, t)] \rho - \frac{\hbar^2}{2m} \left( \frac{\partial^2 \rho}{\partial x_1^2} - \frac{\partial^2 \rho}{\partial x_2^2} \right)$$

$$-i\hbar\gamma(x_1 - x_2)\left(\frac{\partial \rho}{\partial x_1} - \frac{\partial \rho}{\partial x_2}\right) - i\gamma m\omega \coth\left(\frac{\omega\hbar}{2k_B T}\right) (x_1 - x_2)^2 \rho \qquad (37)$$

found out by Caldeira, Cerdeira, and Ramaswamy [10] following the Feynman path integral formalism and making assumptions on the weakness of the damping $\gamma \ll \omega$. Yet our derivation has shown that such an assumption is totally unnecessary to reach (37). Moreover, on the condition that the thermal reservoir is at high temperatures, i.e., $\coth(\omega\hbar/2k_B T) \sim 2k_B T/\omega\hbar$, our master equation (36), along with $\langle \Psi(t) \rangle = 0$ and $I(t) = 1$, yields the master equation

$$i\hbar \frac{\partial \rho}{\partial t} = [V(x_1, t) - V(x_2, t)] \rho - \frac{\hbar^2}{2m} \left( \frac{\partial^2 \rho}{\partial x_1^2} - \frac{\partial^2 \rho}{\partial x_2^2} \right)$$

$$-i\hbar\gamma(x_1 - x_2)\left(\frac{\partial \rho}{\partial x_1} - \frac{\partial \rho}{\partial x_2}\right) - \frac{2i\gamma m k_B T}{\hbar} (x_1 - x_2)^2 \rho \qquad (38)$$

originally found by Caldeira and Leggett [9] upon making use of the path integral techniques and assuming that particle and environment are initially uncorrelated. This assumption is non-realistic and leads to non-physical results [12,16]. Our approach on the contrary has shown that the high-temperature Markovian Caldeira—Leggett equation (38) can be derived for any initial condition as long as averaging and non-Markovian effects could be neglected in our master equation (36) at high temperatures.

In addition, it has been claimed that the Markovian Caldeira—Leggett equation (38) may also lead to unphysical results because it is not of the Lindblad form [3]. To overcome this difficulty, terms has been added to (38) to comply with the Lindblad requirement. Yet, according to our approach such an *ad hoc* procedure is unnecessary given that our quantum master equation (38) could be viewed as the quantization of the non-Markovian Klein—Kramers equation (18) for thermal open systems.

Lastly, because our non-Markovian quantum master equation (36) contains the Markovian Caldeira—Leggett equations (37) and (38) as special cases, we dub it the non-Markovian Caldeira—Leggett equation.

## IV. APPLICATION: THE QUANTUM BROWNIAN FREE MOTION

In order to provide physical significance to the non-Markovian Caldeira—Leggett equation (36), let us consider the correlational function of the form $I(t) = 1 - e^{-\frac{t}{t_c}}$ (see Appendix D) and the case of the quantum Brownian motion of a free particle, $V(x) = 0$, described in quantum phase space by the equation of motion

$$\frac{\partial \overline{W}}{\partial t} = -\frac{p'}{m}\frac{\partial \overline{W}}{\partial x} + 2\gamma \frac{\partial}{\partial p'}\left[p' - \sqrt{\frac{m\mathcal{E}_\hbar(\infty)}{\gamma}}\langle \Psi(t)\rangle\right]\overline{W} + \mathcal{D}_\hbar(t)\frac{\partial^2 \overline{W}}{\partial p'^2}, \quad (39)$$

where the function $\overline{W} \equiv \overline{W}(x, p', t)$ is the Wigner transform

$$\overline{W}(x, p', t) = \frac{1}{2\pi}\int_{-\infty}^{\infty} \rho\left(x + \frac{\eta\hbar}{2}, x - \frac{\eta\hbar}{2}, t\right)e^{-ip'\eta}d\eta, \quad (40)$$

performed upon the equation master (36), and $\mathcal{D}_\hbar(t)$ the time-dependent quantum diffusion coefficient

$$\mathcal{D}_\hbar(t) = 2\gamma m \mathcal{E}_\hbar(\infty)\left(1 - e^{-\frac{t}{t_c}}\right). \quad (41)$$

Upon performing the variable change given by

$$p = p' - \sqrt{\frac{m\mathcal{E}_\hbar(\infty)}{\gamma}}\langle \Psi(t)\rangle \quad (42)$$

so that $\overline{W}(x, p', t) \mapsto W(x, p, t)$, and taking $W(p, t) = \int_{-\infty}^{\infty} W(x, p, t)dx$, Eq. (39) turns out to be rewritten as the quantum Rayleigh equation

$$\frac{\partial W(p, t)}{\partial t} = -\frac{p}{m}\frac{\partial W(p, t)}{\partial x} + 2\gamma \frac{\partial}{\partial p}[pW(p, t)] + \mathcal{D}_\hbar(t)\frac{\partial^2 W(p, t)}{\partial p^2}. \quad (43)$$

To solve (43) we start with the initial condition that couples Brownian particle with the environment

$$W(p, t = 0) = \sqrt{\frac{t_r}{m\pi\hbar}} e^{-\frac{t_r p^2}{m\hbar}}, \qquad (44)$$

where $t_r = (4\gamma)^{-1}$ denotes a sort of relaxation time. The time-dependent solution to (43) reads then

$$W(p, t) = \frac{1}{\sqrt{4\pi A(t)}} e^{-\frac{p^2}{4A(t)}} \qquad (45)$$

with

$$A(t) = \frac{m\hbar}{4t_r} e^{-\frac{t}{t_r}} + \frac{m\omega\hbar}{4} \coth\left(\frac{\omega\hbar}{2k_B T}\right) \left[1 - e^{-\frac{t}{t_r}} + \frac{t_c}{(t_c - t_r)} \left(e^{-\frac{t}{t_r}} - e^{-\frac{t}{t_c}}\right)\right]. \qquad (46)$$

The probability distribution function (45) leads to $\langle P \rangle = 0$ and $\langle P^2 \rangle = 2A(t)$. The average energy is therefore given by

$$\langle E \rangle = \frac{\hbar}{4t_r} e^{-\frac{t}{t_r}} + \frac{\omega\hbar}{4} \coth\left(\frac{\omega\hbar}{2k_B T}\right) \left[1 - e^{-\frac{t}{t_r}} + \frac{t_c}{(t_c - t_r)} \left(e^{-\frac{t}{t_r}} - e^{-\frac{t}{t_c}}\right)\right] \qquad (47)$$

and the momentum fluctuation by

$$\Delta P(t) = \sqrt{\frac{m\hbar}{2t_r} e^{-\frac{t}{t_r}} + \frac{m\omega\hbar}{2} \coth\left(\frac{\omega\hbar}{2k_B T}\right) \left[1 - e^{-\frac{t}{t_r}} + \frac{t_c}{(t_c - t_r)} \left(e^{-\frac{t}{t_r}} - e^{-\frac{t}{t_c}}\right)\right]}. \qquad (48)$$

Because $\Delta P(t) = \Delta P'(t)$ we conclude that the mean value of $\Psi$ in the quantum equation of motion (39) renders unobservable. Moreover, fluctuation (48) gives rise to the following quantum force, $\mathcal{K}(t) = d\Delta P(t)/dt$,

$$\mathcal{K}(t) = -2\gamma\sqrt{2\gamma\hbar m} B(t) \qquad (49)$$

with the dimensionless function $B(t)$ being given by

$$B(t) = \frac{t_r^2 \omega \coth\left(\frac{\omega\hbar}{2k_B T}\right) \left(e^{-\frac{t}{t_r}} - e^{-\frac{t}{t_c}}\right) + (t_c - t_r) e^{-\frac{t}{t_r}}}{\sqrt{(t_c - t_r)\left\{(t_c - t_r) e^{-\frac{t}{t_r}} + t_r \omega \coth\left(\frac{\omega\hbar}{2k_B T}\right) \left[t_r \left(e^{-\frac{t}{t_r}} - 1\right) + t_c \left(1 - e^{-\frac{t}{t_c}}\right)\right]\right\}}}.$$

$$(50)$$

If $B(t) > 0$, then the force (49) is said to be attractive $\mathcal{K}(t) < 0$. On the contrary, if $B(t) < 0$, then (49) renders repulsive, i.e., $\mathcal{K}(t) > 0$. In the steady regime this sort of quantum force vanishes, $\mathcal{K}(\infty) = 0$, whereas at short times $t \ll t_c, t_r$ the force (49) becomes attractive with the constant value $\mathcal{K}(t) = -2\gamma\sqrt{2\gamma\hbar m}$. This fact implies that the quantum force (49) is a non-equilibrium effect in the regime

$0 \leq t < \infty$. Furthermore, the dimensionless factor B($t$) given by (50) is expressed in terms of the following physically accessible time scales: the evolution time $t$ (the observation time, for instance), the correlation time $t_c$, the relaxation time $t_r$, the oscillation time $t_{osc} = \omega^{-1}$, as well as the quantum time $t_q = \hbar/k_B T$.

Taking B($t$)~1 as well as reckoning with the Munro and Gardiner's values for the parameters $\gamma$, $\hbar$, and $m$ in Ref. [12], i.e., $\gamma \sim 10^{11} s^{-1}$, $\hbar \sim 10^{-34} m^2 kg s^{-1}$, and $m \sim 10^{-26}$ kg, we obtain the magnitude $|\mathcal{K}(t)| \sim 10^{-13}$ N. This simple numerical example suggests that the strength of the quantum thermal force (49) could be measured in experiences, for instance, using trapped ions [17] in which measurement of forces of order of yoctonewton, i.e., $10^{-24}$ N, has been reported.

At zero temperature, i.e., $\coth(\omega\hbar/2k_B T) \sim 1$, the dimensionless factor (50) becomes

$$B^{(T=0)}(t) = \frac{t_r^2 \omega \left(e^{-\frac{t}{t_r}} - e^{-\frac{t}{t_c}}\right) + (t_c - t_r)e^{-\frac{t}{t_r}}}{\sqrt{(t_c - t_r)\left\{(t_c - t_r)e^{-\frac{t}{t_r}} + \omega t_r \left[t_r\left(e^{-\frac{t}{t_r}} - 1\right) + t_c\left(1 - e^{-\frac{t}{t_c}}\right)\right]\right\}}}, \quad (52)$$

while at high temperatures, i.e., as $\coth(\omega\hbar/2k_B T) \sim 2k_B T/\omega\hbar$, it turns out to be

$$B(t) = \frac{2k_B T t_r^2 \left(e^{-\frac{t}{t_r}} - e^{-\frac{t}{t_c}}\right) + \hbar(t_c - t_r)e^{-\frac{t}{t_r}}}{\sqrt{\hbar(t_c - t_r)\left\{\hbar(t_c - t_r)e^{-\frac{t}{t_r}} + 2k_B T t_r \left[t_r\left(e^{-\frac{t}{t_r}} - 1\right) + t_c\left(1 - e^{-\frac{t}{t_c}}\right)\right]\right\}}}. \quad (53)$$

In the classical limit, $\hbar \to 0$, the sort of quantum force (49) becomes $F(t) = -2\gamma\sqrt{mk_B T}G(t)$ with the dimensionless function G($t$) being

$$G(t) = \frac{t_r \left(e^{-\frac{t}{t_r}} - e^{-\frac{t}{t_c}}\right)}{\sqrt{(t_c - t_r)\left[t_r\left(e^{-\frac{t}{t_r}} - 1\right) + t_c\left(1 - e^{-\frac{t}{t_c}}\right)\right]}}. \quad (54)$$

Taking into account G($t$)~1, $\gamma \sim 10^{11} s^{-1}$, $m \sim 10^{-26}$ kg, $k_B \sim 10^{-23} m^2 kg s^{-2} K^{-1}$, and $T \sim 1000$K, we find $|F(t)| \sim 10^{-22}$ N. That is the magnitude of the classical thermal force exerted by a heat bath at 1000K upon the Brownian motion of a free particle of mass $10^{-26}$ kg.

At short times the dimensionless function (54) approximates to G($t$)~ $-(2\gamma t_c)^{-1/2}$. Hence the non-equilibrium classical thermal force reads

$$F(t) = \sqrt{\frac{2\gamma m k_B T}{t_c}} \tag{55}$$

which blows up in the Markovian limit $t_c \to 0$. This result reveals the pivotal importance of non-Markovian effects for the existence of the physical concept of non-equilibrium thermal force (54) in the classical realm. In mathematical parlance, non-Markovian features account for the differentiability property of the root mean square momentum $\Delta P(t) = \sqrt{\langle P^2 \rangle - \langle P \rangle^2}$ in the classical domain.

## V. SUMMARY AND OUTLOOK

In summary, our article bears the following novel results:

a) The first upshot is that non-Markovian effects upon the Brownian motion can turn up in the classical realm through the generalized Klein—Kramers equation (19);

b) Our second finding is that we reach the non-Markovian generalization of the Caldeira—Leggett equation [Eq.(36)] without resorting to path-integral approach to open quantum systems;

c) Lastly, on the ground of the non-Markovian Caldeira—Leggett quantum master equation (8) we have put forward the existence of the quantum thermal force far from equilibrium [Eq.(49) with (50)] that is valid as much at zero temperature as at high temperatures.

We hope our concept of non-equilibrium quantum thermal force could be experimentally borne out, thereby refreshing interest in the Caldeira—Leggett equation for describing quantum dynamics of open systems in the Gaussian approximation. Moreover, we reckon our dynamical-quantization approach to quantum open systems may shed some light upon the understanding of physical mechanisms underlying the interaction between a quantum object (e.g. a qubit) and its environment in so as to minimize the destructive influence of decoherence process, for instance.

In conclusion, in a forthcoming paper we intend to look at open quantum systems described by our non-Gaussian non-Markovian quantum master equation (B3).

## Acknowledgments

I thank Professor Maria Carolina Nemes for the scientific support and FAPEMIG (Fundação de Amparo à Pesquisa do Estado de Minas Gerais) for the financial support under the contract CEX-00103/10. Referee's constructive criticisms will be acknowledged, as well.

# Appendix A: The non-Markovian Klein—Kramers equation

The Langevin equations (3) and (4) give rise to the Kolmogorov equation [18]

$$\frac{\partial \mathcal{F}(x,p,t)}{\partial t} = \mathbb{K}\mathcal{F}(x,p,t), \tag{A1}$$

where the Kolmogorovian operator $\mathbb{K}$ acts upon the probability distribution function $\mathcal{F}(x,p,t)$ according to

$$\mathbb{K}\mathcal{F}(x,p,t) = \sum_{k=1}^{\infty}\sum_{r=0}^{k} \frac{(-1)^k}{r!\,(k-r)!} \frac{\partial^k}{\partial x^{k-r}\partial p^r}\left[A^{(k-r,r)}(x,p,t)\mathcal{F}(x,p,t)\right] \tag{A2}$$

with

$$A^{(k-r,r)}(x,p,t) = \lim_{\varepsilon\to 0}\left[\frac{\langle(\Delta X)^{k-r}\rangle\langle(\Delta P)^r\rangle}{\varepsilon}\right]. \tag{A3}$$

The increments $\Delta X = X(t+\varepsilon) - X(t)$ and $\Delta P = P(t+\varepsilon) - P(t)$ in the coefficients (A3) are calculated from (3) and (4) on the basis of

$$\Delta X = \frac{1}{m}\int_t^{t+\varepsilon} P(t)dt, \tag{A4}$$

$$\Delta P = -\frac{dV(X)}{dX}\varepsilon - 2\gamma\int_t^{t+\varepsilon} P(t)dt + b\int_t^{t+\varepsilon}\Psi(t)dt. \tag{A5}$$

The average values, $\langle(\Delta X)^{k-r}\rangle\langle(\Delta P)^r\rangle$, in turn are to be calculated about the sharp values $q$ and $p'$, i.e.,

$$\mathcal{F}_{XP\Psi}(x,p,\psi,t) = \delta(x-q)\delta(p-p')\mathcal{F}_\Psi(\psi,t). \tag{A6}$$

The Kolmogorov equation (A1) describes the time evolution of a Brownian particle immersed in a general non-Gaussian environment. According to the Pawula theorem [19], there exists no non-Gaussian approximation to (A1) complying with the positivity of $\mathcal{F}(x,p,t)$. A sufficient condition leading to a Gaussian approximation to (A1) is then to consider

$$|x_2 - x_1|^3 \ll 0, \tag{A7}$$

such that the non-Gaussian coefficients in (A3) can vanish. In (A7), $x_2 = x(t+\varepsilon)$ and $x_1 = x(t)$ are associated with the increment $\Delta X = X(t+\varepsilon) - X(t)$. So, in the Gaussian approximation (A7) the Kolmogorov equation (A1) changes into the Fokker—Planck equation

$$\frac{\partial \mathcal{F}}{\partial t} = -\frac{\partial}{\partial x}[A^{(1,0)}\mathcal{F}] - \frac{\partial}{\partial p}[A^{(0,1)}\mathcal{F}] + \frac{1}{2}\frac{\partial^2}{\partial x^2}[A^{(2,0)}\mathcal{F}] + \frac{\partial^2}{\partial x \partial p}[A^{(1,1)}\mathcal{F}]$$
$$+ \frac{1}{2}\frac{\partial^2}{\partial p^2}[A^{(0,2)}\mathcal{F}], \tag{A8}$$

whereby $\mathcal{F} = \mathcal{F}(x, p, t)$, the drift coefficients are given by

$$A^{(1,0)}(x, p, t) = \lim_{\varepsilon \to 0}\left[\frac{\langle \Delta X \rangle}{\varepsilon}\right] = \frac{p}{m}, \tag{A9}$$

and

$$A^{(0,1)}(x, p, t) = \lim_{\varepsilon \to 0}\left[\frac{\langle \Delta P \rangle}{\varepsilon}\right] = -2\gamma p - \frac{dV(x)}{dx} + b\langle \Psi(t) \rangle, \tag{A10}$$

whereas the diffusion coefficients are

$$A^{(2,0)}(x, p, t) = \lim_{\varepsilon \to 0}\left[\frac{\langle (\Delta X)^2 \rangle}{\varepsilon}\right] = 0, \tag{A11}$$

$$A^{(1,1)}(x, p, t) = \lim_{\varepsilon \to 0}\left[\frac{\langle \Delta X \rangle \langle \Delta P \rangle}{\varepsilon}\right] = 0, \tag{A12}$$

$$A^{(0,2)}(x, p, t) = \lim_{\varepsilon \to 0}\left[\frac{\langle (\Delta P)^2 \rangle}{\varepsilon}\right] = b^2 I(t). \tag{A13}$$

Taking into account (A9—A13), the Fokker—Planck equation (A8) turns out to written as

$$\frac{\partial \mathcal{F}}{\partial t} = -\frac{p}{m}\frac{\partial}{\partial x}\mathcal{F} + \frac{\partial}{\partial p}\left[\frac{dV(x)}{dx} + 2\gamma p - b\langle \Psi(t) \rangle\right]\mathcal{F} + 2\gamma m \mathcal{E}(t)\frac{\partial^2}{\partial p^2}\mathcal{F} \tag{A14}$$

with the diffusion energy

$$\mathcal{E}(t) = \frac{b^2}{4\gamma m} I(t). \tag{A15}$$

In (A14), the mean value of $\Psi(t)$ reads

$$\langle \Psi(t) \rangle = \lim_{\varepsilon \to 0}\frac{1}{\varepsilon}\int_{t}^{t+\varepsilon} \langle \Psi(t') \rangle dt', \tag{A16}$$

while the dimensionless time-dependent function $I(t)$ in (A15) is given by

$$I(t) = \lim_{\varepsilon \to 0}\frac{1}{\varepsilon}\iint_{t}^{t+\varepsilon} \langle \Psi(t')\Psi(t'') \rangle dt' dt''. \tag{A17}$$

## Appendix B: Quantizing the Kolmogorov equation

We write down the Kolmogorov equation (A1) in the form

$$\frac{\partial \mathcal{F}}{\partial t} = -\frac{p}{m}\frac{\partial}{\partial x}\mathcal{F} + \frac{\partial}{\partial p}\left[\frac{dV(x)}{dx} + 2\gamma p - b\langle\Psi(t)\rangle\right]\mathcal{F} + 2\gamma m\mathcal{E}(t)\frac{\partial^2}{\partial p^2}\mathcal{F} + \overline{\mathbb{K}}\mathcal{F}, \quad (B1)$$

where the non-Gaussian terms are given by

$$\overline{\mathbb{K}}\mathcal{F} = \sum_{k=3}^{\infty}\sum_{r=0}^{k}\frac{(-1)^k}{r!(k-r)!}\frac{\partial^k}{\partial x^{k-r}\partial p^r}\left[A^{(k-r,r)}(x,p,t)\mathcal{F}(x,p,t)\right] \quad (B2)$$

with the coefficients (A3).

After performing the Fourier transformation (24) and making use of the quantization conditions (26) and (27), we obtain the non-Gaussian quantum master equation

$$i\hbar\frac{\partial\rho}{\partial t} = \left[\mathcal{V}_{\text{eff}}^{(\hbar)}(x_1,t) - \mathcal{V}_{\text{eff}}^{(\hbar)}(x_2,t)\right]\rho - \frac{\hbar^2}{2m}\left(\frac{\partial^2\rho}{\partial x_1^2} - \frac{\partial^2\rho}{\partial x_2^2}\right) - i\hbar\gamma(x_1-x_2)\left(\frac{\partial\rho}{\partial x_1} - \frac{\partial\rho}{\partial x_2}\right)$$
$$- \frac{i\mathcal{D}_\hbar(t)}{\hbar}(x_1-x_2)^2\rho + \mathcal{O}\rho \quad (B3)$$

with the non-Gaussian contributions given by

$$\mathcal{O}\rho = \int_{-\infty}^{\infty}\overline{\mathbb{K}}\mathcal{F}\left(\frac{x_1-x_2}{2},p,t\right)e^{ip\frac{(x_1-x_2)}{\hbar}}dp. \quad (B4)$$

In the Gaussian approximation (A7), i.e., $|x_1 - x_2|^3 \ll 1$, the non-Gaussian terms (B4) do vanish. Accordingly, the non-Gaussian quantum master equation (B3) reduces to our Gaussian master equation (29).

## Appendix C: The physics of the environment

Let us assume that the environment can be imagined as a heat bath comprising of a set of $N$ quantum harmonic oscillators having the same oscillation frequency $\omega$ in thermal equilibrium at temperature $T$. According to the quantum statistical thermodynamics [20], the internal energy $U$ of this harmonic quantum heat bath is given by $U = N\bar{\epsilon}$, where

$$\bar{\epsilon} = \frac{\omega\hbar}{2}\left(\frac{e^{\frac{\omega\hbar}{k_BT}} + 1}{e^{\frac{\omega\hbar}{k_BT}} - 1}\right) = \frac{\omega\hbar}{2}\coth\left(\frac{\omega\hbar}{2k_BT}\right) \tag{C1}$$

is the mean energy of one oscillator. So, after identifying the steady quantum diffusion energy $\mathcal{E}_\hbar(\infty)$ of the Brownian particle described by Eq. (32) with the thermal mean energy $\bar{\epsilon}$ (C1) of the quantum heat bath, we obtain

$$\mathcal{E}_\hbar(\infty) = \bar{\epsilon} = \frac{\omega\hbar}{2}\coth\left(\frac{\omega\hbar}{2k_BT}\right), \tag{C2}$$

where the quantum energy, $\mathcal{E}_\hbar^{(T=0)}(\infty) = \omega\hbar/2$, corresponds to the zero-point thermal diffusion energy inherent in the heat bath at zero temperature, and $\mathcal{E}(\infty) = k_BT$ the classical thermal diffusion energy of the quantum heat bath at high temperatures $T \gg \omega\hbar/2k_B$.

Furthermore, it is worth remarking that in the quantized expressions (C1) and (C2), the oscillation frequency $\omega$, the Boltzmann constant $k_B$, the temperature $T$, as well as the friction constant $\gamma$ are not quantized quantities in the sense that they do not rely upon the Planck constant $\hbar$ in the quantum realm.

## Appendix D: The correlational function

The non-Markovian correlation function $I(t) = 1 - e^{-\frac{t}{t_c}}$ can be obtained by taking the autocorrelation function

$$\langle \Psi(t')\Psi(t'') \rangle = \left(1 - e^{-\frac{(t'+t'')}{2t_c}}\right)\delta(t' - t'')$$

into Eq. (13). The correlation time $t_c$ denotes the time during which the fluctuations of the Langevin stochastic force $L(t) = b\gamma\Psi(t)$ remain correlated at times $t'$ and $t''$. Physically, the correlation time is roughly viewed as the duration of a collision of the environmental particles with the Brownian particle.

# References

[1] M. A. Nielsen and I. L. Chuang, *Quantum Computation and Quantum Information* (Cambridge University Press, Cambridge, 2000); *Quantum Computation and Quantum Information Theory*, edited by C. Macchiavello, G. M. Palma, and A. Zeilinger (World Scientific, Singapore, 2000); *The Physics of Quantum Computation,* edited by D. Bouwmeester, A. Ekert, and A. Zeilinger (Springer, Berlin, 2000); S. Stenholm and K-A. Suominen, *Quantum Approach to Informatics* (J. Wiley, New Jersey, 2005); *Manipulating Quantum Coherence in Solid State Systems*, edited by M. E. Flatte and E. Tifrea (Springer, Dordrecht, 2007); L. C. L. Hollenberg *et al.*, Phys. Rev. A **64**, 042309 (2001).

[2] J. G. Peixoto de Faria and M. C. Nemes, J. Phys.: Math. Gen. **31**, 7095 (1998); R. C. de Berredo *et al.*, Physica Scripta **57**, 533 (1998); D. Kohen, C. C. Marston, and D. J. Tannor, J. Chem. Phys. **107**, 5236 (1997).

[3] G. Lindblad, Commun. Math. Phys. 48, 119 (1976).

[4] A. O. Bolivar, *Quantum-Classical Correspondence: Dynamical Quantization and the Classical Limit* (Springer, Berlin, 2004).

[5] H.-P. Breuer and F. Petruccione, *The Theory of Open Quantum Systems* (Oxford University Press, Oxford, 2002).

[6] H. J. Carmichael, *Statistical Methods in Quantum Optics* 1 (Springer, Berlin, 1999); C. W. Gardiner and P. Zoller, *Quantum Noise* (Springer, Berlin, 2004).

[7] U. Weiss, *Quantum Dissipative Systems* (World Scientific, Singapore, 2008).

[8] S. Gao, Phys. Rev. Lett. **79**, 3101 (1997); **80**, 5703 (1998); **82**, 3377 (1998); B. Vacchini, Phys. Rev. Lett. **84**, 1374 (2000); **87**, 028902 (2000); G. W. Ford, R. F. O'Connell, Phys. Rev. Lett. **82**, 3376 (1999); R. F. O'Connell, Phys. Rev. Lett. **87**, 028901 (2001); H. M. Wiseman, W. J. Munro, Phys. Rev. Lett. **80**,5702 (1998).

[9] A.O. Caldeira and A. Leggett, Physica A **121**, 587 (1983).

[10] A.O. Caldeira, H.A. Cerdeira, and R. Ramaswamy, Phys. Rev. A **40**, 3438 (1989).

[11] M. Schlosshauer, *Decoherence and the Quantum-to-Classical Transition* (Springer, Berlin, 2007); E. Joss, H. D. Zeh, C. Kiefer, D. Giulini, J. Kupsch, and L.-O. Stamatescu, *Decoherence and the Appearance of a classical World in Quantum Theory* (Springer, Berlin, 2003).

[12] W. J. Munro and C. W. Gardiner, Phys. Rev. A **53**, 2633 (1996); V. Ambegaokar, Ber. Bunsenges. Phys. Chem. **95** (1991) 400; L. Diòsi, Europhys. Lett. **22**, 1(1993); Physica (Amsterdam) **199A**, 517 (1993); S. Gao, Phys. Rev. B **57**, 4509 (1998); S. M. Barnett, J. Jeffers, and J. D. Cresser, J. Phys.: Condens. Matter **18** (2006) S401; A. Tameshtit and J. E. Sipe, Phys. Rev. Lett. **77**, 2600 (1996); R. Karrlein and H. Grabert, Phys. Rev. E **55**, 153 (1997).

[13] A.O. Bolivar, Phys. Rev. A **58**, 4330 (1998); Random Oper. and Stoch. Equ. **9**, 275 (2001); Physica A (Amsterdam) **301**, 219 (2001); Phys. Lett. A **307**, 229 (2003); Can. J. Phys. **81**, 663 (2003); Phys. Rev. Lett. **94**, 026807 (2005); M. Razavy, *Classical and Quantum Dissipative Systems* (Imperial College Press, London, 2005).

[14] H. Risken, *The Fokker-Planck Equation: Methods of Solution and Application* (Springer-Verlag, Berlin, 1989) 2nd ed; W. T. Coffey, Y. P. Kalmykov, and J. T. Waldron, *The Langevin Equation: with Applications to Stochastic Problems in Physics, Chemistry and Electrical Engineering* (World Scientific, Singapore, 2004) 2nd ed; N. G. van Kampen, *Stochastic Processes in Physics and Chemistry* (Elsevier, Amsterdam, 2007) 3rd ed; Gardiner, C. W.: *Handbook of Stochastic Methods: for Physics, Chemistry, and the Natural Sciences* (Springer, Berlin (2004) 3rd ed; R. M. Mazo, *Brownian Motion: Fluctuations, Dynamics and Applications* (Oxford University Press, New York, 2002).

[15] O. Klein, Arkiv f. Math. Astr. Phys. **16**, 1 (1922); H. A. Kramers, *Physica* **7**, 284 (1940).

[16] C. Presilla, R. Onofrio, and M. Patriarca, J. Phys. A: Math. Gen. **30**, 7385 (1997); A. Tameshitit and J.E. Sipe, Phys. Rev. Lett. **77**, 2600 (1996); M. Patriarca, Nuovo Cimento **B 111**, 61 (1996); C. Morais Smith and A. O. Caldeira, Phys. Rev. A **36**, 3509 (1987); C. Morais Smith and A. O. Caldeira, Phys. Rev. A **41**, 3103 (1990); V. Hakim and V. Ambegaokar, Phys. Rev. A **32**, 423 (1985); H. Grabert, P. Schramm, G-L. Ingold, Phys. Rep. **168**, 115 (1988); N. G. van Kampen, J. Stat. Phys. **115**, 1057 (2004).

[17] M. J. Biercuk, H. Uys, and J. W. Britton, e-print quant-ph/1004.0780.

[18] A. Kolmogorov, Math. Ann. **104**, (1931) 414.

[19] R. F. Pawula, Phys. Rev. **162**, 186 (1967).

[20] R. Tolman, *The Principles of Statistical Mechanics* (Dover, New York, 1979).